\documentstyle[preprint,aps,prl]{revtex}
\draft
\include{psfig}
\begin{document}
\title{Aging as dynamics in configuration space}
\author{ Walter Kob $^{+}$, Francesco Sciortino $^{o}$ and Piero Tartaglia $^{o}$} 
\address{ $^{+}$ Institute of Physics, Johannes-Gutenberg University,
D-55099 Mainz, Germany}
\address{ $^{o}$ Dipartimento di Fisica and Istituto Nazionale
per la Fisica della Materia, Universit\'a di Roma {\it La Sapienza},
P.le Aldo Moro 2, I-00185 Roma, Italy.}

\maketitle
\begin{abstract}
The relaxation dynamics of many
disordered systems, such as structural glasses, proteins, granular
materials or spin glasses, is not completely frozen even at very low
temperatures~\cite{glassy_systems1,angell94}. This residual
motion leads to a change of the properties of the material, a process
commonly called aging. Despite recent advances in the theoretical
description of such aging processes, the microscopic mechanisms leading
to the aging dynamics are still a matter of dispute~\cite{bouchaud98,cugliandolo,parisi}.
In this letter we investigate the aging dynamics of a simple glass
former by means of molecular dynamics computer simulation. Using the
concept of the inherent structure we give evidence that aging dynamics
can be understood as a decrease of the effective configurational 
temperature $T$  of the system. 
From our results we conclude that the equilibration process is 
faster when the system is quenched to $T_c$, the critical
$T$  of mode-coupling theory~\cite{mct1}, and that thermodynamic
concepts~\cite{cugliandolo97,nieuwenhuizen98} are useful to describe
the out-of-equilibrium aging process.
\end{abstract}

Despite their constitutive differences, many complex disordered
materials show a strikingly similar dynamical
behavior\cite{glassy_systems1,glassy_systems3}. In such systems, the
characteristic relaxation time increases by many decades upon a small
variation of the external control parameters such as $T$  or
density. Correlation functions  show power-law and stretched
exponential behavior as opposed to a simple exponential decay.  Also
the equilibration process is frequently of non-exponential nature, and
is often so slow to give rise to strong out-of-equilibrium phenomena,
commonly named aging. It has recently been recognized that the
similarity in the equilibrium dynamics for many systems 
might be related to a similarity
in the structure of their configuration space (often called
energy-landscape)~\cite{stillinger95} and that this structure can be
studied best in the out-of-equilibrium situation. In this
letter we investigate the energy-landscape of an aging system
and show that by means of this landscape it is indeed possible to
establish a close connection between the equilibrium and 
out-of-equilibrium properties of the same system.

For the case of glass-forming liquids, whose characteristic relaxation
time increases by more than 13 decades when $T$ is
decreased by a modest amount, the configuration space is given by the 
$3N$ dimensional space spanned by the spatial coordinates of the
$N$ atoms. In 1985, Stillinger and Weber introduced the concept of
the inherent structure (IS), which can be defined as
follows~\cite{stillinger95}: for any configuration of particles the IS
is given by that point which is reached by a steepest descent procedure
in the potential energy if one uses the particle configuration as
starting point of the minimization, i.e. the IS is the location of
the nearest local potential-energy minimum in configuration space. 
By this method configuration space
can be decomposed in a unique way into the basins of attraction of
all IS of the systems and the time-evolution of a
system in configuration space can be described as a
progressive exploration of different IS. Here we
determine the properties of the IS in equilibrium as well as in the 
out-of-equilibrium situation. From the comparison of the IS in these
different situations we elucidate the
dynamics of the system during the aging and learn about
the structure of configuration space and the physics of the
glassy dynamics.

The system we study is a binary mixture of Lennard-Jones
particles whose equilibrium dynamics has been investigated in great
detail~\cite{kob_lj1,kob_lj2,sastry98}. It has been found that
this dynamics can be well described by the so-called
mode-coupling theory\cite{mct1} with a critical temperature 
$T_c=0.435$. To study the non-equilibrium dynamics we quench at time
zero the equilibrated system from an initial temperature $T_i=5.0$ to a
final temperature $T_f \in\{0.1, 0.2, 0.3, 0.4, 0.435\}$. In
Fig.~\ref{fig1} we show $e_{\rm IS}$, the average energy per particle
in the IS, as a function of $T$ (equilibrium case --- panel a)
and as a function of time (out-of-equilibrium case --- panel b),
respectively.  In agreement with the results of Ref.~\cite{sastry98} we
find that in equilibrium $e_{\rm IS}$ is almost independent of
$T$ for $T \geq 1.0$, i.e.  when the thermal energy $k_{\rm B}
T$ is larger than the depth of the Lennard-Jones pair potential. At lower
$T$, $e_{\rm IS}$ shows a significant $T$ dependence
confirming that on decreasing $T$ the
system is resident in deeper minima. The 
relation $e_{\rm IS}(T)$ can be inverted, $T=T(e_{\rm IS})$, and we propose
to use this relation to associate {\it in the non-equilibrium case} to
each value of $e_{\rm IS}(t)$ an effective temperature $T_e(e_{\rm
IS}(t))$ [see Fig.~\ref{fig1}]\cite{previouswork}.  
By associating an equilibrium $T$
value to an $e_{\rm IS}(t)$ value we can describe the
(out-of-equilibrium) time dependence of $e_{\rm IS}$ during the aging
process as a progressive exploration of configuration space valleys
with lower and lower energy or, equivalently, as a progressive
thermalization of the configurational potential energy. We find that,
for all studied $T_f$, the equilibration process is composed by three
regimes (Fig.~\ref{fig1}b).  An early-time regime, during which the
equilibrating system explores basins with  high IS energy and in 
which $e_{\rm IS}(t)$ shows little $t$ dependence.
This regime 
is followed by one at intermediate-time in which $e_{\rm
IS}(t)$ decreases with a power-law with an exponent $0.13\pm 0.02$, 
independent of $T_f$. This scale-free 
$t$-dependence is evidence that the aging process is  
a self-similar process. At even longer $t$ a third
regime is observed for the lowest $T_f$, characterized by a slower
thermalization rate.  The cross-over between these three time regimes
is controlled by the value of $T_f$ .

We show next that during the equilibration process the system visits
the same type of minima as the one visited in equilibrium. 
We evaluate the curvature of the potential energy at the IS
as a function of $T$  (for the equilibrium case) and as a
function of time $t$ (for the out-of-equilibrium case) by calculating the
density of states $P(\nu)$, i.e. the distribution of 
normal modes with frequency $\nu$.  Before we discuss the frequency
dependence of $P(\nu)$ we first look at its first moment, $\bar{\nu}$,
a quantity which can be calculated with higher accuracy than the
distribution itself. The $T$-dependence of $\bar{\nu}$ in
equilibrium and its $t$-dependence in non-equilibrium are shown in
Figs.~\ref{fig2}a and \ref{fig2}b, respectively. 
Note that Fig.~\ref{fig1} and Fig.~\ref{fig2} are quite similar. This
demonstrates 
that during the progressive thermalization, the aging system explores  IS with
the same $e_{\rm IS}$ and with the same curvature as the one visited in
equilibrium, i.e. the same type of minima.  
The similarity of Fig.~\ref{fig1} and Fig.~\ref{fig2} 
also demonstrates that $T_e(e_{\rm IS})$ can indeed serve as a
$T$  which characterizes the typical configuration occupied by
the system.
We also consider the
frequency dependence of $P(\nu)$, which is shown for the equilibrium
case in Fig.~\ref{fig3}.  In the inset we show $P(\nu)$ at the highest
and lowest $T$  investigated and we find that the
dependence of $P(\nu)$ on $T$  is rather weak. To better visualize the
weak $T$-dependence of $P(\nu)$
we discuss in the following
$P(\nu)-P_0(\nu)$, where $P_0(\nu)$ is the (equilibrium) distribution
at $T=0.446$, the lowest $T$  at which we were able to
equilibrate the system. In Fig.~\ref{fig3} we show  $P(\nu)-P_0(\nu)$ and
from it we see that the main effect of a change in $T$  is that
with decreasing $T$  the modes at high $\nu$ disappear and
that more modes in the vicinity of the peak appear. We also find that 
if an analogous plot is made for the out-of-equilibrium data the same
pattern is observed, i.e. that with increasing $t$ $P(\nu)$ becomes
narrower and more peaked.

We next show that $T_e(e_{\rm IS})$ completely
determines $P(\nu)$ during the aging process.
For this we see from Fig.~\ref{fig1} that 
$T_e=0.6$ corresponds to $t \approx 1600$ for
$T_f=0.435$, $0.4$, and $0.3$, and to $t \approx 4000$ and $t \approx 25000$ 
for
$T_f=0.2$ and $0.1$, respectively (see dashed lines in Fig.~\ref{fig1}).
If $T_e$ has thermodynamic meaning the {\it non-equilibrium}
$P(\nu)$ at these times should be the same as
the {\it equilibrium} $P(\nu)$  at $T=0.6$. 
These functions are plotted
in Fig.~\ref{fig4} (curves with filled symbols). We
find that the different distribution functions 
are superimposed,
demonstrating the validity of the proposed interpretation of $T_e$ as
effective $T$ . That this collapse of the curves is not a
coincidence can be recognized by the second set of curves which is
shown in   Fig.~\ref{fig4} (curves with open symbols). These curves correspond
to $T_e=0.5$ for which the corresponding times from
Fig.~\ref{fig1} are $t \approx 16000$ 
for $T_f=0.435$, and $0.4$, and the $t \approx 25000$ for 
$T_f=0.3$~\cite{comment}. Also for $T_e=0.5$ the
different $P(\nu)$  can be considered to be identical
within the accuracy of the data. Note that the two set of curves
corresponding to the two values of $T_e$ are clearly different, 
showing that our data has a sufficiently high precision to distinguish
also values of $T_e$ which are quite close together. From the present
data we conclude that minima with the same value of $e_{\rm IS}$
do indeed have the same distribution of curvature.

To discuss the results of this paper it is useful to recall
insight gained from the analysis of instantaneous normal modes
(INM) of supercooled liquids\cite{keyes_review}. The INM studies have
demonstrated that the slowing down of the dynamics in supercooled
liquids is accompanied by a progressive decrease in the number of
so-called double-well modes, i.e. the number of directions in
configuration space where the potential energy surface has a saddle
leading to a new minimum, a condition stronger than the local
concavity. It has been demonstrated~\cite{sciortino98} that the number
of double-well modes vanishes on approach to $T_c$. 
Thus for $T > T_c$ the system is always
located on a potential energy landscape which, in at least one direction,
is concave (i.e. the system sits on a saddle point) whereas at
$T < T_c$ the system is located in the vicinity of the local
minima, i.e. the landscape is convex. This result can be rephrased
by saying that $T_c$ is the $T$  at which the
thermal energy  $k_B T_c$ is of the same order as the height of the
lowest-lying saddle point above the nearest IS, i.e. above $e_{\rm
IS}$.  The energy difference between $k_B T +e_{\rm IS}$ and the
lowest-lying saddle point energy can be chosen as an effective
($T$-dependent) barrier height. This observation, which should hold
true also for the Lennard-Jones system studied here, and the results
presented here lead us to the following view of the
energy landscape and the aging process.  For high $T$, $k_BT$
is significantly higher than the lowest lying saddle energy and the
effective barriers between two minima are basically zero, i.e. the
system can  explore the whole configuration space. At 
$T \approx 1.0$ the system starts to become trapped in that part
of configuration space which has a value of $e_{\rm IS}$ which is less
than the one at high $T$ (Fig.~\ref{fig1} and
Ref.~\cite{sastry98}) and the properties of the IS start to become
relevant.  This point of view of the structure of phase space can now
be used to understand the aging dynamics. At the beginning of the
quench the system is still in the large part of configuration space
which corresponds to (high) $T_i$.  Although $k_BT_f$
is now relatively small, the effective barriers are still zero and the
system can move around unhindered and it  moves to minima which
have a lower energy. The rate of this exploration is related to the
number of double-well directions accessible within $k_BT_f$, which
explains why in Fig.~\ref{fig1}b the curves with small $T_f$ stay at
the beginning longer on the plateau than the ones with larger $T_f$.
With increasing $t$ the system starts to find IS which have a lower
and lower energy and  $e_{\rm IS}(t)$ starts to decrease.  Note
that apart from the $T_f$ dependence of the rate of exploration just
discussed, this search seems to be independent of $T_f$, since in
Fig.~\ref{fig1}b the slope of the curves at intermediate times does not
depend on $T_f$. 

With increasing $t$ the system finds IS with lower and lower energies
and  decreases its $T_e$. From the above discussion on the INM we
know that with decreasing $T$ the height of the
effective barriers also increases and it can be expected that the search of
the system becomes inefficient once it has reached a
$T_e$  at which the energy difference between the lowest-lying
saddle and $e_{\rm IS}$ becomes of the order of $k_{\rm B} T_f$.
Therefore we expect that once this stage has been reached the 
$t$-dependence of $e_{\rm IS}$ will change and this is what we find, 
as shown  in
Fig.~\ref{fig1}b in the curves for $T_f=0.2$ and 0.1 at $t \approx 
10^4$. We also note that the $T_e$ at which this crossover occurs will
increase with decreasing $T_f$, in agreement with the result shown in
Fig.~\ref{fig1}. For times larger than this crossover $t$ the system
no longer explores the configuration space by moving along unstable
modes but rather by means of a hopping mechanism in which barriers are
surmounted. This hopping mechanism, although not  efficient for moving
the system through configuration space, still allows the system to
decrease its configurational energy and its 
$T_e$ further. Thus the cross-over from the self-similar
process to the activated dynamics, which in equilibrium is located
close to $T_c$, is in the non-equilibrium case $T_f$ dependent.
We conclude that in order to obtain configurations which are
relaxed as well as possible (within a given time span) one should
quench the system to $T_c$ 
in order to exploit as much as possible the low-lying saddle points.

The presented picture implies that, if hopping
processes were not present, $T_e$  for the
system would always be above 
$T_c$.  Although hopping processes
are always present, they might be so inefficient that even for long times the
value of $T_e$ is above $T_c$. From Fig.~\ref{fig1} we recognize that
this is the case for the present study. We note that 
theoretical mean-field predictions derived for 
$p-$spin models\cite{andrea} and recent
extensions of
the ideal MCT equations to non-equilibrium processes~\cite{latz99}
conclude that system quenched below $T_c$  
always remain in that part of configuration
space corresponding to $T>T_c$.

Finally we stress that the present analysis of the aging process 
strongly support the use of the
IS as an appropriate tool for the identification of an effective $T$ 
at which the configurational potential energy is at equilibrium. This
opens the way for detailed comparisons with recent out-of-equilibrium
thermodynamics approaches\cite{nieuwenhuizen98} and for detailed
estimates of the configurational entropy.

\begin{figure}[h]
\caption{a) $e_{\rm IS}$, the potential energy of the inherent
structure in equilibrium as a function of temperature (panel a) and as
a function of time during the aging process (panel b). The dashed line
is used to define the effective temperature $T_e(t)$ in the
non-equilibrium case.  The microscopic model we consider is a binary
(80:20) mixture of Lennard-Jones particles, which in the following we
will call type $A$ and type $B$ particles. 
The interaction between two particles of type $\alpha$ and $\beta$,
with $\alpha,\beta \in \{\rm A,B\}$, is given by $V_{\alpha\beta}=
4\epsilon_{\alpha\beta} [(\sigma_{\alpha\beta}/r)^{12}-
(\sigma_{\alpha\beta}/r)^6]$. The parameters $\epsilon_{\alpha\beta}$
and $\sigma_{\alpha\beta}$ are given by $\epsilon_{AA}=1.0$,
$\sigma_{AA}=1.0$, $\epsilon_{AB}=1.5$, $\sigma_{AB}=0.8$,
$\epsilon_{BB}=0.5$, and $\sigma_{BB}=0.88$.  The potential is
truncated and shifted at $r_{\rm cut}=2.5\sigma_{\alpha\beta}$.
$\sigma_{\rm AA}$ and $\epsilon_{\rm AA}$ are chosen as the unit of
length and energy, respectively (setting the Boltzmann constant $k_{\rm
B}=1.0$). Time is measured in units of $\sqrt{ m \sigma_{\rm AA}^2
/48\epsilon_{\rm AA}}$, where $m$ is the mass of the particles. 1000
particles were used at a fixed density of 1.2. Further information on
the equilibrium and non-equilibrium simulations can be found in Refs.
\protect\cite{kob_lj1,kob_lj2,lj_age} By using copies
of the system at different times $t$ since the quench, we calculated
the IS of the system by means of a conjugate gradient method. To
improve the statistics of the results we averaged them over 8-10
independent runs.  The same method was also used to determine the IS
for the system at equilibrium in the range $5.0 \geq T \geq
0.446$.}
\label{fig1}
\end{figure}

\begin{figure}[h]
\caption{a) Temperature dependence of $\bar{\nu}$, the first moment
of the density of states, in equilibrium.
b) Time dependence of $\bar{\nu}$ during the aging process.
}
\label{fig2}
\end{figure}

\begin{figure}[h]
\caption{Frequency dependence in equilibrium of $P(\nu)$, the density
of states at frequency $\nu$.  Main figure: Temperature dependence of
$P(\nu)$ for all temperatures investigated. In order to see this
dependence clearer we have subtracted from these distributions
$P_0(\nu)$, the equilibrium distribution function at $T=0.446$.
Inset: Comparison of $P(\nu)$ at $T=5.0$ and $T=0.446$.
}
\label{fig3}
\end{figure}

\begin{figure}[h]
\caption{Comparison of $P(\nu)$ in the out-of-equilibrium situation
at different values of $T_f$ and different times (but at the same
$T_e$) with $P(\nu)$ at equilibrium with $T=T_e$. Filled symbols and
bold dashed line: $T_e=T=0.5$. Open symbols and bold solid line:
$T_e=T=0.6$.
}
\label{fig4}
\end{figure}

\setcounter{figure}{0}

\eject

\begin{figure}
\centerline{\psfig{figure=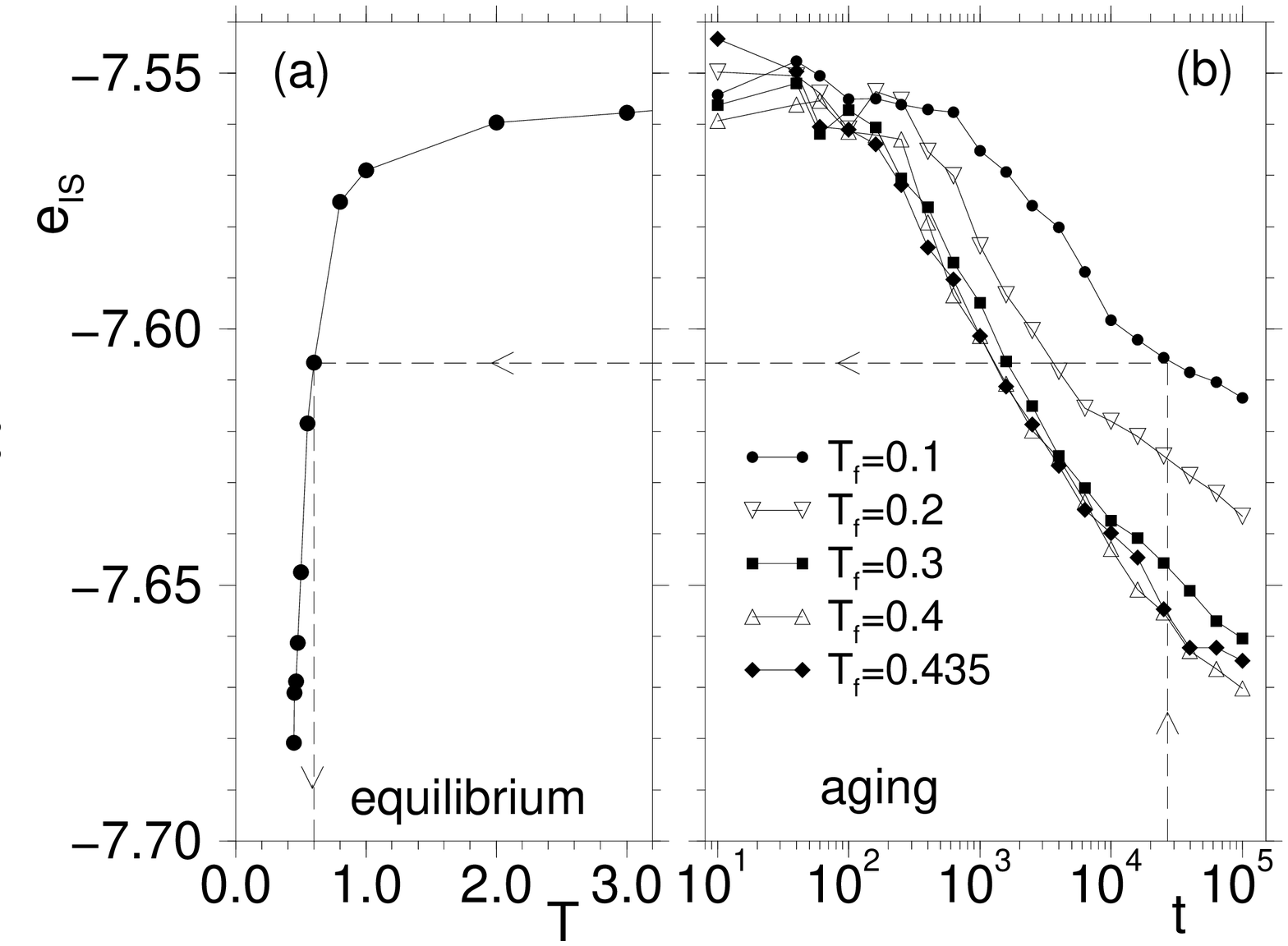,height=13cm,width=16cm,clip=,angle=0.}}
\caption{W. Kob, F. Sciortino and P. Tartaglia}
\end{figure}

\newpage

\begin{figure}
\centerline{\psfig{figure=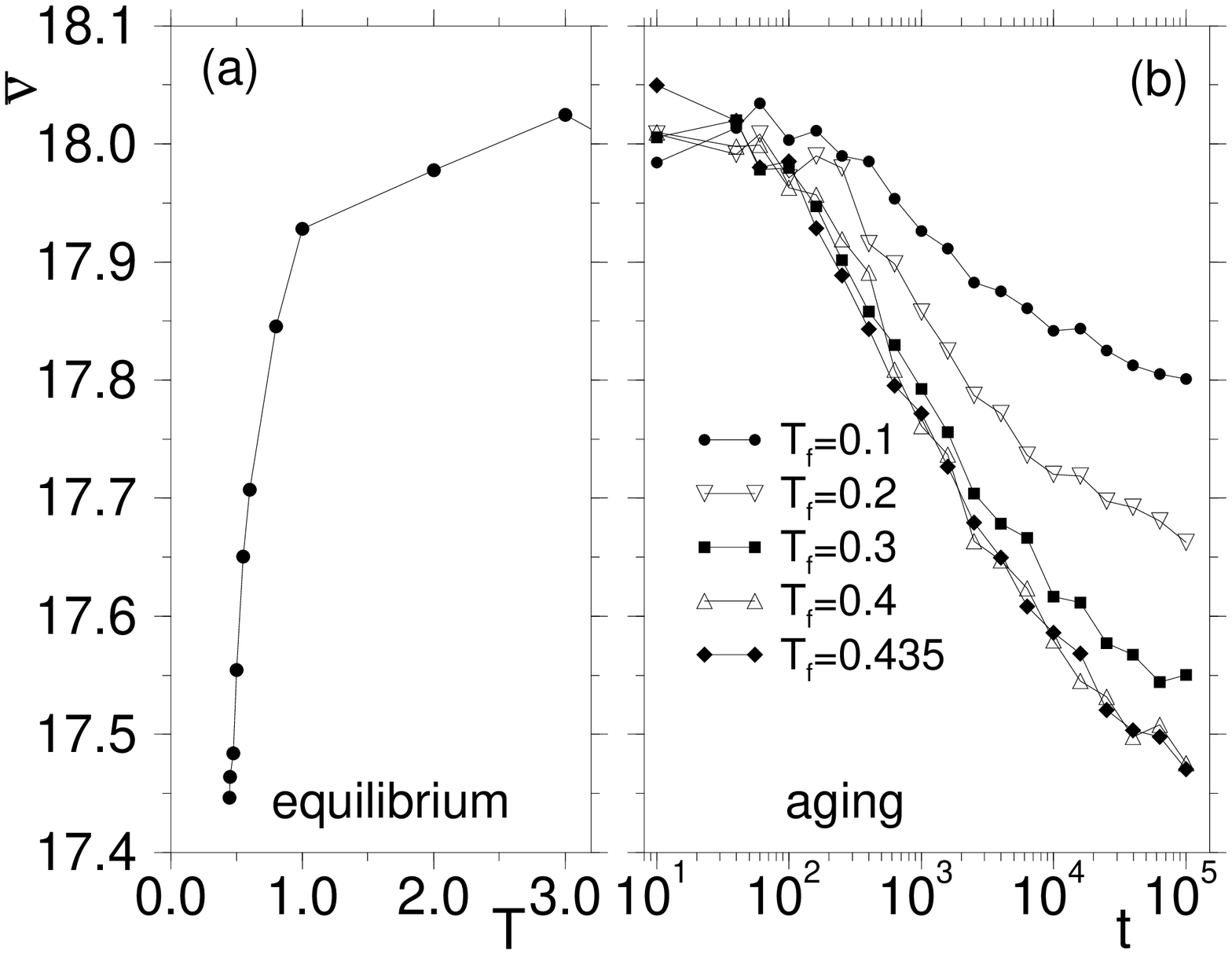,height=13cm,width=16cm,clip=,angle=0.}}
\caption{W. Kob, F. Sciortino and P. Tartaglia}
\end{figure}

\newpage

\begin{figure}
\centerline{\psfig{figure=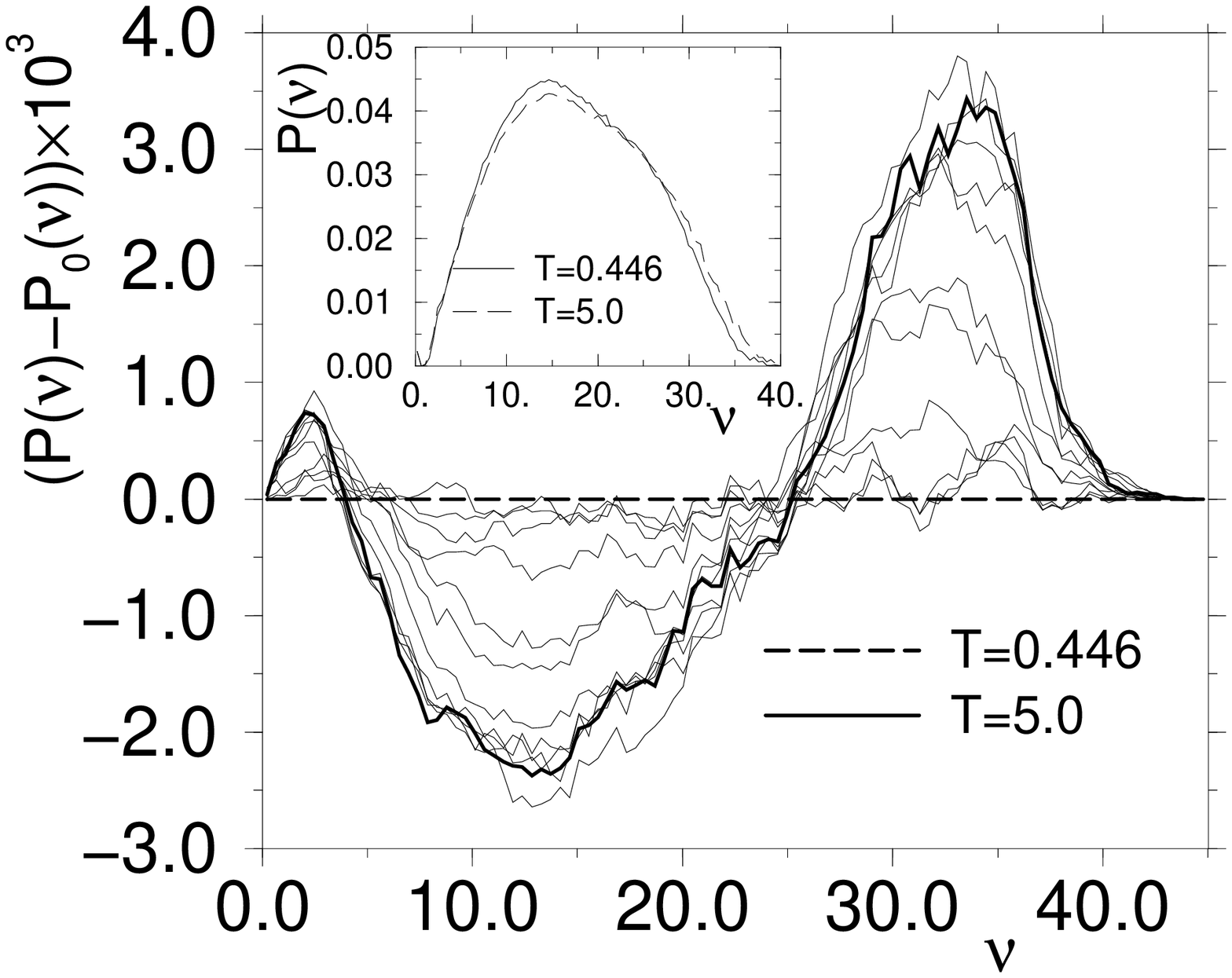,height=13cm,width=16cm,clip=,angle=0.}}
\caption{W. Kob, F. Sciortino and P. Tartaglia}
\end{figure}

\newpage

\begin{figure}
\centerline{\psfig{figure=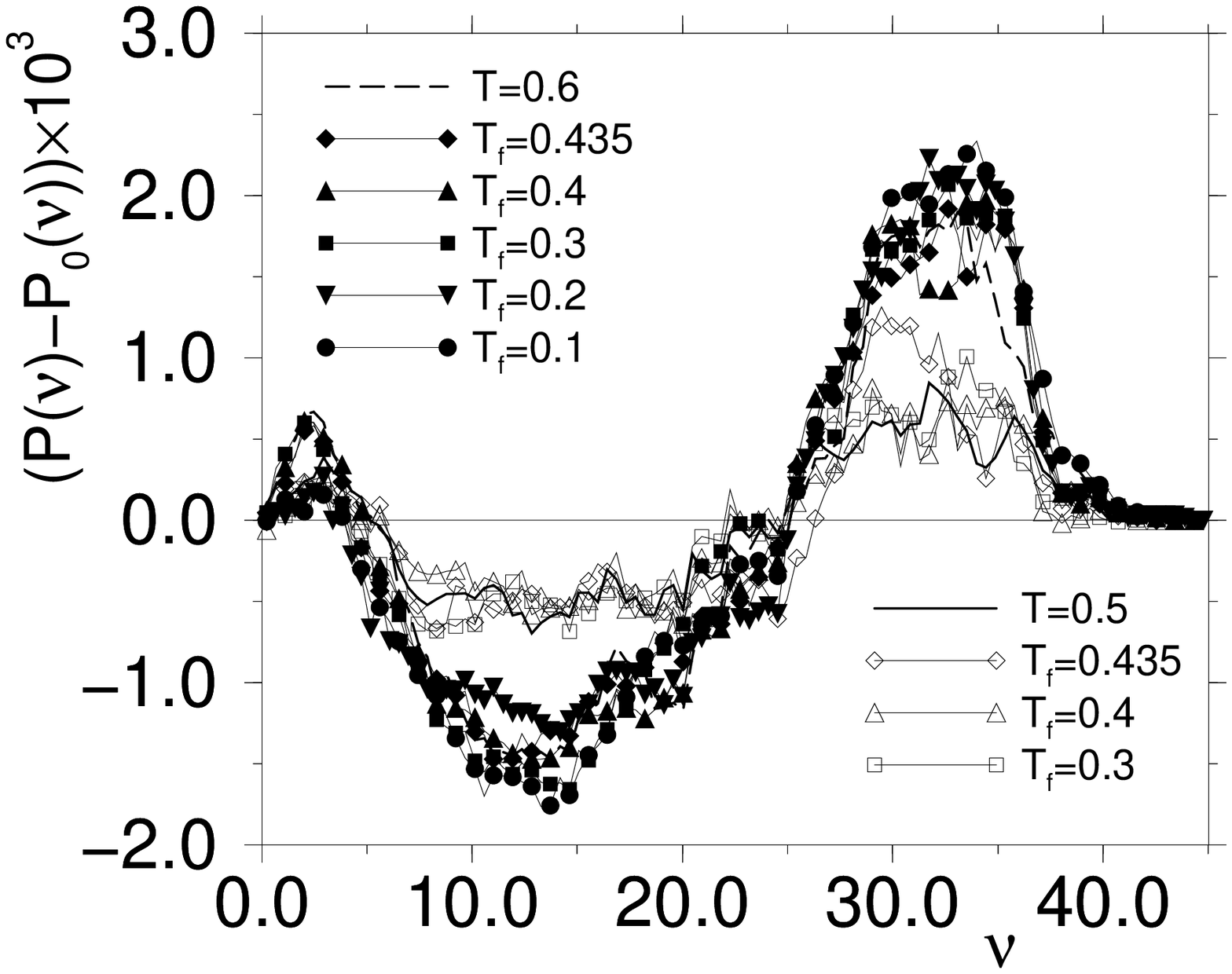,height=13cm,width=16cm,clip=,angle=0.}}
\caption{W. Kob, F. Sciortino and P. Tartaglia}
\end{figure}


\begin{references}
\bibitem{glassy_systems1} 
See, e.g., the articles in {\it Complex
Behavior of Glassy Systems} Eds.: M. Rubi and C. Perez-Vicente
(Springer, Berlin, 1997).

\bibitem{angell94}
C. A. Angell, P. H. Poole, and J. Shao,
Il Nuovo Cimento {\bf 16D}, 993-1025 (1994).

\bibitem{bouchaud98}
J.-P. Bouchaud, L. F.  Cugliandolo, J. Kurchan, and M. M\'ezard,
p. 161-223 in {\it Spin Glasses and Random Fields}, Ed.: A.P.
Young (World Scientific, Singapore, 1998).

\bibitem{cugliandolo} 
L. F. Cugliandolo, J. Kurchan, and P. Le Doussal 
Phys. Rev. Letts. {\bf 76} 2390  (1996).

\bibitem{parisi} 
G. Parisi, 
Phys. Rev. Letts. {\bf 79}  3660 (1997).


\bibitem{mct1}
W. G\"otze,
J. Phys.: Condens. Matter {\bf 11}, A1-A45 (1999).

\bibitem{cugliandolo97}
L. F. Cugliandolo, J. Kurchan, and L. Peliti,
Phys. Rev. E {\bf 55}, 3898-3914 (1997).

\bibitem{nieuwenhuizen98} 
Th. M. Nieuwenhuizen,  
Phys. Rev. Letts. {\bf 80} 5580 (1998);
preprint cond-mat/9807161, and references therein.

\bibitem{glassy_systems3}
H. Frauenfelder, {\it et al.} {\it Landscape Paradigms in Physics and
Biology} Special issue in Phyica D {\bf 107} 1997.

\bibitem{stillinger95}
F. H. Stillinger,
Science, {\bf 267} 1935-1939 (1995).

\bibitem{kob_lj1}
W. Kob and H. C. Andersen,
Phys. Rev. Lett. {\bf 73}, 1376--1379 (1994).

\bibitem{kob_lj2}
W. Kob and H. C. Andersen,
Phys. Rev. E {\bf 51}, 4626--4641 (1995).
W. Kob and H. C. Andersen,
Phys. Rev. E {\bf 52}, 4134--4153 (1995).

\bibitem{sastry98}
S. Sastry, P. G. Debenedetti, and F. H. Stillinger,
Nature {\bf 393}, 554-557 (1998).

\bibitem{lj_age}
W. Kob and J.-L. Barrat,
Phys. Rev. Lett. {\bf 78}, 4581--4584 (1997).


\bibitem{previouswork} The concept of effective temperature in
glassy system has a long history 
(See for example A.Q. Tool, J. Am. Ceram.
Soc. {\bf 29} 240 (1946)).
We refer the interested reader to articles quoted in Ref.
\protect\cite{nieuwenhuizen98} 


\bibitem{comment}
For this value of $T_e$ no time can be read off for $T_f=0.1$, and 0.2
since for this value of $T_f$ the effective temperature is higher than
0.5 in the time range of our simulation.

\bibitem{keyes_review}
T. Keyes,
J. Phys. Chem. A {\bf 101}, 2921-2930 (1997).

\bibitem{sciortino98}
F. Sciortino and P. Tartaglia,
Phys. Rev. Lett. {\bf 78}, 2385 (1997). 
See also for mean field models J. Kurchan and L. Laloux,
{\it Phase space geometry and slow dynamics}
 cond-mat/9510079. 

\bibitem{comment}
For this value of $T_e$ no time can be read off for $T_f=0.1$, and 0.2
since for this value of $T_f$ the effective temperature is higher than
0.5 in the time range of our simulation.

\bibitem{andrea} A. C. Crisanti and  H.J. Sommers 
  J. Phys. I (France) 5, 805 (1995).

\bibitem{latz99}
A. Latz,  unpublished.


\end{references}
\end{document}